\documentclass[11pt]{jaa}
\usepackage{natbib}
\usepackage{rotating}
\usepackage{subcaption}
\usepackage{wrapfig}

\usepackage[]{hyperref}
\hypersetup{colorlinks = true, linkcolor = blue, anchorcolor = red, citecolor = blue, filecolor = red, urlcolor = red, hypertexnames=true}

\begin{document}\sloppy

\title{MCMC Optimisation applied to Double Stars from Miller \& Pitman}


\author{Maksym Ersteniuk\textsuperscript{1}, Timothy Banks\textsuperscript{1,2,*}, Edwin Budding\textsuperscript{3,4}, and Michael. D. Rhodes\textsuperscript{5}}
\affilOne{\textsuperscript{1}Dept. of Physical Sci. \& Engineering, Harper College, 1200 W Algonquin Rd, Palatine, IL 60067, USA\\}
\affilTwo{\textsuperscript{2}Nielsen, 675 6th Ave, New York 10011, USA\\}
\affilThree{\textsuperscript{3}Carter Observatory, 40 Salamanca Road, Kelburn, Wellington 6012, New Zealand\\}
\affilFour{\textsuperscript{4}School of Chemical \& Physical Sciences, Victoria University of Wellington, PO Box 600, Wellington 6140, NZ\\}
\affilFive{\textsuperscript{5}Brigham Young University, Provo, Utah 84602, USA}


\twocolumn[{

\maketitle

\corres{tim.banks@nielsen.com}

\msinfo{1 January 2024}{1 January 2024}

\begin{abstract}
Model orbits have been fitted to 27 physical double stars listed in a 1922 catalogue. A Markov Chain Monte Carlo technique was applied to estimate best fitting values and associated uncertainties for the orbital parameters.  Dynamical masses were calculated using parallaxes from the {\em Hipparcos} mission, and are presented in this paper together with the estimates of the orbital parameters for the 27 systems.  The resulting mass estimates of the current study are in good agreement with a recently published study, as are comparisons with the orbital parameters listed by the Washington Double Star catalog, confirming the validity of the optimisation methodology.
\end{abstract}


}]


\doinum{12.3456/s78910-011-012-3}
\artcitid{\#\#\#\#}
\volnum{000}
\year{2024}
\pgrange{1--}
\setcounter{page}{1}
\lp{8}


\section{Introduction} \label{sec:intro}

Double stars make popular observing targets for various reasons including an interest in the practicalities of obtaining valuable scientific data with small telescopes.  Such measurements can potentially indicate whether the two stars, observed to be in close proximity on the sky are a physical double, as the two stars should slowly shift their relative positions with time as they orbit about each other. Binary stars are important to astronomy as they allow directly determining stellar masses.  Where the stars are not observed to be following such orbits,  their proximity in the sky might mean that the stars are actually gravitationally remote from each other and therefore simply in a similar line of sight from Earth (i.e., an optical double). In practice, it turns out that this situation is often not the case. In passing we note that \cite{Argyle_2004}, \cite{MacEvoy_2015} and \cite{Argyle_2019} provide useful background materials for observing and analysing visual doubles.

A major goal of the current paper is to outline the testing of an algorithm based on Markov Chain Monte Carlo (MCMC) optimization. This note documents our final testing of a Bayesian-based methodology through comparison on systems with known results, placing these results into the literature for later use by the double star community. The rationale behind these tests is that agreement of our findings with literature results would lend confidence for later general applications of the method, such as for systems without known orbital solutions. A noteworthy point is that our method provides uncertainties for the derived parameters, something not provided for many orbital solutions in the literature.

The paper therefore outlines the automated estimation of values and uncertainties of orbital parameters to a selection of physical double binaries listed in \cite{Miller_1922}, and in particular from their Table 1 of `First Class' systems that those authors considered to possess well determined orbital estimates (and therefore good systems for the planned testing).  \citeauthor{Miller_1922} did not present the orbital solutions and so we make use of the parameter estimates adopted in the Washington Double Star (WDS) catalogue \citep{Mason_2022}. We sourced the positional data from the WDS, current to 2023. Our paper presents estimates for the orbital parameters for these systems using all the available data. We will show that these results were found to be in good agreement with the solutions given in the WDS with the advantage that single sigma uncertainties are presented for all the estimated parameters.  This is reached through the use of an optimisation technique based on Bayesian statistics, which is described below. 


\begin{figure}
    \centering 
 \begin{subfigure}{0.45\textwidth}
  \includegraphics[width=\linewidth]{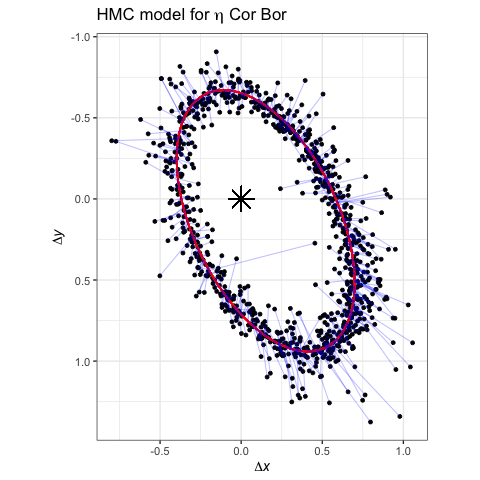}
  \caption{$\eta$ Cor Bor}
  \label{fig:eta_cor_bor_orbit}
 \end{subfigure} \\
\begin{subfigure}{0.45\textwidth}
  \includegraphics[width=\linewidth]{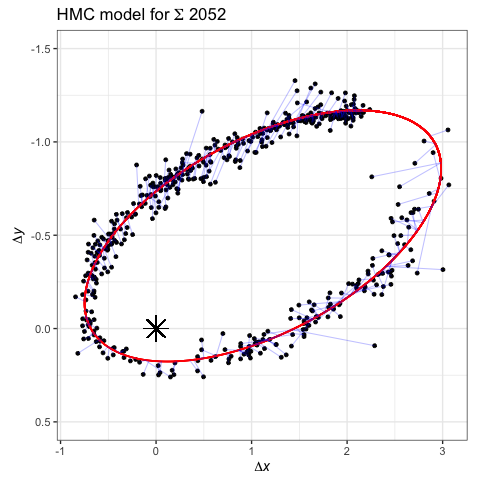}
  \caption{$\Sigma \: 2052$}
  \label{fig:sigma_2052_orbit}
\end{subfigure}\hfil 
 \caption{Observations and model orbits are shown for two representative systems from those analysed in this paper. North is upwards and east increases to the right, as is the convention in many visual binary papers.  The red curve plots the model orbit, while observations (the position of the secondary star relative to the primary star at a given point in time) are shown as black dots. Blue lines connect the observations to their predicted positions given the derived orbital parameters. The star symbol at $(\Delta x, \Delta y) \sim (0,0)$ is the location of the primary star in each system.
\label{fig:orbit_plots}}
\end{figure}


\section{MCMC} \label{sec:mcmc}
\noindent
A Markov Chain is a probabilistic model describing the likelihood of possible future states based on the currently observed state.  It is a `memory-less' process, typically based on a matrix giving the transition probabilities between one observed state to another.  The current state of the process depends only on the immediately previous one.  A chain is built up of repeated steps through this transition matrix.  

MCMC combines such chains with a Monte Carlo approach, or basically random probabilities \citep{Privault_2013, Hogg_2018}.  This combination allows MCMC to explore and then characterize a distribution by randomly sampling that distribution without requiring knowledge of the distribution's mathematical properties \citep{van_Ravenzwaaji_2018}. It is a Bayesian statistical technique, where inferences are made on unknown quantities (such as model parameters or predictions) by combining prior `knowledge' (often called `beliefs' in the literature) about those quantities together with observations.  

A technique called NUTS (No U-Turn Sampler) avoids the random walk behaviour of more simple MCMC algorithms, making a faster exploration of possible model parameter sets and a faster convergence to an optimal set of parameter estimates \citep{Hoffman_2014}. It handles multiple parameter models better than more simple techniques, which struggle with these higher dimensional problems.  We made use of this Hamiltonian Monte Carlo (HMC) technique for these reasons.  

We are not the first authors to apply a MCMC method to visual double star data, but the technique is not yet widely used in the field (see, e.g., \citealt{Sahlman_2013}, \citealt{Lucy_2014}, and \citealt{Mendez_2017}). 


\section{Analysis} \label{sec:analysis}

\noindent
The orbit of a (visual) binary star system can be described on the $xy$ plane as (see, e.g. \citealt{Ribas_2002}):
$$
        x = \frac{a ( 1 - e^2 )}{1 + e \: cos{\nu}} \big [ \cos{ (\nu + \omega)} \sin{\Omega} + sin{(\nu + \omega)} \cos{\Omega} \cos{i} \big ]
$$
$$
    y = \frac{a ( 1 - e^2 )}{1 + e \: cos{\nu}} \big [ \cos{(\nu + \omega)} \cos{\Omega} - sin{(\nu + \omega)} \sin{\Omega} \cos{i} \big ]
$$

\noindent
$a$ is the semi-major axis of the orbit, measured in arc-seconds.  $e$ is the orbital eccentricity. $\nu$ is the true anomaly (or function of time) of the orbit of the stars about their barycenter. $i$ is the inclination, the angle between the plane of projection and the orbital plane.  Position angles were precessed to the year 2000.

\begin{sidewaystable*}
    \caption{{\bf Parameter estimates from MCMC fitting} to the Miller \& Pitman systems. Uncertainties are single sigma (one standard deviation). See the text for the explanation of the symbols used as the column titles other than `Epoch', which is the time of phase zero for the orbital ephemeris, the orbital period ($P$) in years, and `WDS', which gives the Washington Double Star catalogue ID.  The final column (G) on the right employs the orbit grading scheme of \cite{Worley_1983} as outlined in \cite{Hartkopf_2001}.}
    \centering
    {\small 
    \begin{tabular}{l|r|r|r|r|r|r|r|l|l}
    \hline
System              & $P$                & $a$                 & $e$               & $\omega$          & $i$             & $\Omega$        & Epoch               & WDS              &   G \\
    \hline
42 Com  Ber         & $ 26.48 \pm 5.34$  & $0.6671 \pm 0.1105$ & $0.237 \pm 0.154$ & $  12.4 \pm 4.3$  &$ 89.9 \pm 3.0$  & $317.0 \pm 97.9$& $2011.65 \pm 6.35$  & J13100+1732AB    &   3 \\ 
70 Oph              & $ 88.29 \pm 0.04$  & $4.5920 \pm 0.0087$ & $0.516 \pm 0.003$ & $ 122.7 \pm 0.2$  &$ 58.5 \pm 0.1$  & $165.6 \pm 0.4$ & $1896.00 \pm 0.07$  & J18055+0230AB    &   1 \\ 
85 Pegasi           & $ 26.29 \pm 0.03$  & $0.8112 \pm 0.0174$ & $0.388 \pm 0.029$ & $ 292.9 \pm 1.9$  &$130.1 \pm 1.5$  & $258.6 \pm 3.8$ & $1963.62 \pm 0.27$  & J00022+2705AB    &   1 \\ 
99 Herc             & $ 56.01 \pm 0.18$  & $1.0513 \pm 0.0716$ & $0.751 \pm 0.037$ & $ 216.4 \pm 5.8$  &$142.0 \pm 3.5$  & $ 61.3 \pm 5.8$ & $1997.44 \pm 0.28$  & J18070+3034Aa,Ab &   2 \\ 
A 88                & $ 12.10 \pm 0.07$  & $0.1439 \pm 0.0085$ & $0.262 \pm 0.089$ & $170.1 \pm 5.00$  &$ 57.4 \pm 4.2$  & $280.4 \pm 28.1$& $2007.30 \pm 0.94$  & J18384-0312AB    &   2 \\ 
$\beta$ 80          & $ 97.07 \pm 0.51$  & $0.6707 \pm 0.0248$ & $0.733 \pm 0.028$ & $ 231.9 \pm 74.8$ &$178.0 \pm 12.0$ & $127.7 \pm 74.8$& $1905.29 \pm 0.42$  & J23189+0524AB    &   4 \\ 
$\beta$  524        & $ 31.68 \pm 0.24$  & $0.2286 \pm 0.0133$ & $0.769 \pm 0.028$ & $  26.5 \pm 1.5$  &$ 59.0 \pm 1.6$  & $ 94.2 \pm 1.4$ & $1996.54 \pm 0.13$  & J02537+3820      &   3 \\ 
$\beta$ 612         & $ 22.39 \pm 0.03$  & $0.2137 \pm 0.0060$ & $0.614 \pm 0.035$ & $  39.6 \pm 2.8$  &$137.1 \pm 2.6$  & $  5.5 \pm 3.6$ & $1952.46 \pm 0.15$  & J13396+1045      &   2 \\ 
$\beta$  1111       & $ 39.60 \pm 0.21$  & $0.2360 \pm 0.0082$ & $0.290 \pm 0.054$ & $  44.1 \pm 5.2$  &$137.8 \pm 3.9$  & $214.9 \pm 11.9$& $1918.86 \pm 1.16$  & J14234+0827      &   2 \\ 
Castor              & $523.75 \pm 18.23$ & $7.2337 \pm 0.1492$ & $0.401 \pm 0.016$ & $  40.2 \pm 0.3$  &$ 65.8 \pm 0.2$  & $116.6 \pm 2.0$ & $1953.67 \pm 1.40$  & J07346+3153AB    &   3 \\ 
$\epsilon$ Equulei  & $102.60 \pm 0.48$  & $0.6476 \pm 0.0208$ & $0.760 \pm 0.038$ & $ 105.9 \pm 0.9$  &$ 87.4 \pm 1.0$  & $ 14.8 \pm 2.8$ & $1921.42 \pm 0.55$  & J20591+0418AB    &   3 \\ 
$\eta$  Cass        & $482.07 \pm 8.26$  &$11.9964 \pm 0.0991$ & $0.497 \pm 0.009$ & $ 278.7 \pm 0.8$  &$144.9 \pm 0.4$  & $ 91.0 \pm 1.3$ & $1890.20 \pm 0.60$  & J00491+5749A     &   3 \\ 
$\eta$  Cor Bor     &$  41.53 \pm 0.07$  &$ 0.8765 \pm 0.0054$ & $0.270 \pm 0.010$ & $ 203.6 \pm 0.5$  &$121.7 \pm 0.4$  & $321.8 \pm 2.1$ & $1975.49 \pm 0.23$  & J15232+3017      &   1 \\ 
$\gamma$  Cor Bor   &$  93.54 \pm 0.69$  &$ 0.7469 \pm 0.0142$ & $0.506 \pm 0.026$ & $ 111.5 \pm 0.6$  &$ 85.0 \pm 0.7$  & $244.3 \pm 2.5$ & $1934.36 \pm 0.70$  & J15427+2618AB    &   2 \\ 
$\kappa$ Pegasi     &$  11.54 \pm 0.06$  &$ 0.2260 \pm 0.0078$ & $0.332 \pm 0.059$ & $ 287.3 \pm 2.0$  &$ 70.5 \pm 2.3$  & $225.2 \pm 11.5$& $1991.19 \pm 0.36$  & J21446+2539AB    &   2 \\ 
Krueger 60          &$  44.68 \pm 0.06$  &$ 2.3707 \pm 0.0296$ & $0.417 \pm 0.015$ & $ 153.2 \pm 20.8$ &$  9.1 \pm 4.8$  & $ 96.8 \pm 20.7$& $1970.27 \pm 0.23$  & J22280+5742A     &   2 \\ 
$\mu$  Herc         & $ 42.95 \pm 0.06$  &$ 1.3406 \pm 0.0118$ & $0.157 \pm 0.017$ & $  61.4 \pm 0.6$  &$113.0 \pm 0.6$  & $192.3 \pm 6.2$ & $2007.66 \pm 0.73$  & J17465+2743Aa,Ab &   1 \\ 
Procyon             & $ 40.78 \pm 0.10$  & $4.2782 \pm 0.0918$ & $0.406 \pm 0.025$ & $ 100.2 \pm 5.0$  &$150.1 \pm 2.8$  & $272.9 \pm 5.9$ & $1927.04 \pm 0.46$  & J07393+0514A     &   2 \\ 
$\Sigma$ 518        & $225.74 \pm 7.73$  & $6.9767 \pm 0.2314$ & $0.434 \pm 0.050$ &  $151.8 \pm 0.6$  & $73.1 \pm 0.9$  & $ 48.1 \pm 6.1$ & $1844.82 \pm 1.87$  & J04153-0739      &   3 \\ 
$\Sigma$ 1938 Aa,Ab & $  3.75 \pm 0.01$  & $0.1043 \pm 0.0049$ & $0.289 \pm 0.042$ & $ 127.2 \pm 1.5$  & $52.2 \pm 1.6$  & $310.1 \pm 5.9$ & $2006.44 \pm 0.07$  & J15245+3723      &   1 \\ 
$\Sigma$ 1938 Ba,Bb & $267.31 \pm 4.06$  & $1.4535 \pm 0.0119$ & $0.590 \pm 0.011$ & $1 72.7 \pm 1.0$  & $45.0 \pm 0.90$ & $ 20.5 \pm 2.1$ & $1866.20 \pm 0.85$  & J15245+3723      &   2 \\ 
$\Sigma$ 2052       & $237.71 \pm 6.56$  & $2.3485 \pm 0.0430$ & $0.774 \pm 0.010$ & $  93.3 \pm 0.4$  & $72.3 \pm 0.3$  & $230.4 \pm 0.7$ & $1921.12 \pm 0.25$  & J16289+1825AB    &   2 \\
$\Sigma$ 2107       & $266.85 \pm 6.60$  & $1.0081 \pm 0.0146$ & $0.555 \pm 0.019$ & $  50.7 \pm 3.5$  &$150.1 \pm 1.8$  & $116.5 \pm 4.5$ & $1895.46 \pm 1.06$  & J16518+2840AB    &   2 \\ 
$\Sigma$ 2173       & $ 46.42 \pm 0.12$  & $0.9809 \pm 0.0069$ & $0.180 \pm 0.014$ & $ 151.8 \pm 0.4$  &$ 80.3 \pm 0.4$  & $ 30.6 \pm 4.6$ & $1962.85 \pm 0.59$  & J17304-0104AB    &   1 \\ 
Sirius              & $ 50.09 \pm 0.02$  & $7.6762 \pm 0.0530$ & $0.600 \pm 0.007$ & $  23.6 \pm 0.4$  & $44.3 \pm 0.4$  & $213.4 \pm 0.7$ & $1994.29 \pm 0.08$  & J06451-1643A     &   2 \\ 
$\tau$  Cygni       & $ 49.23 \pm 0.60$  & $0.8970 \pm 0.0113$ & $0.234 \pm 0.021$ & $ 158.6 \pm 1.42$ & $45.6 \pm 1.0$  & $239.9 \pm 5.1$ & $1940.21 \pm 0.57$  & J21148+3803A     &   2 \\ 
$\zeta$  Herc       & $ 34.45 \pm 0.03$  & $1.3517 \pm 0.0123$ & $0.458 \pm 0.011$ & $  50.0 \pm 0.9$  & $48.4 \pm 0.7$  & $247.1 \pm 1.4$ & $1933.42 \pm 0.12$  & J16413+3136AB    &   1 \\ 
    \hline 
    \end{tabular} }
    \label{tab:system_parameters}
\end{sidewaystable*}

\begin{figure*}
    \centering 
 \begin{subfigure}{0.45\textwidth}
  \includegraphics[width=\linewidth]{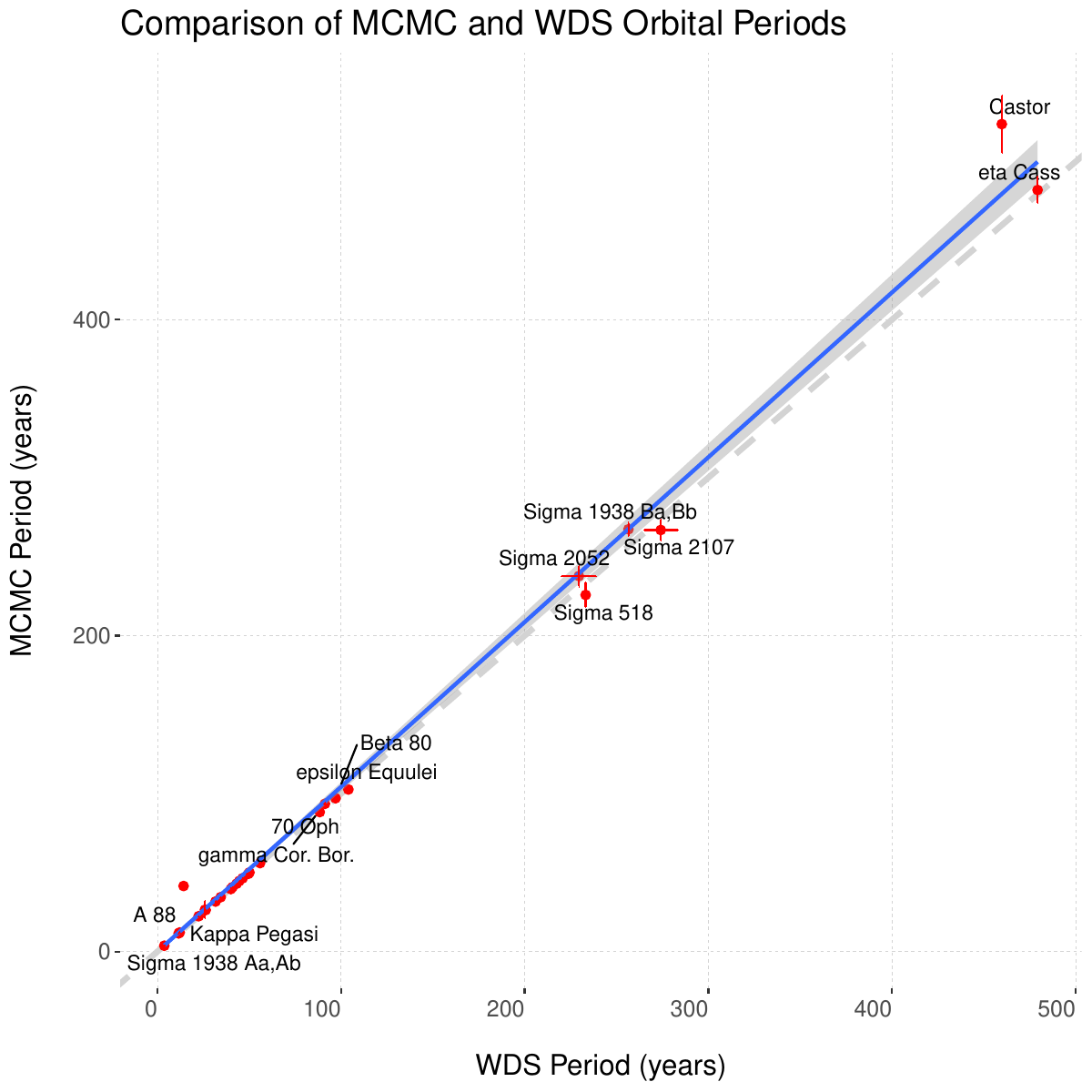}
  \caption{$P$}
  \label{fig:Period_comp}
 \end{subfigure}
\begin{subfigure}{0.45\textwidth}
  \includegraphics[width=\linewidth]{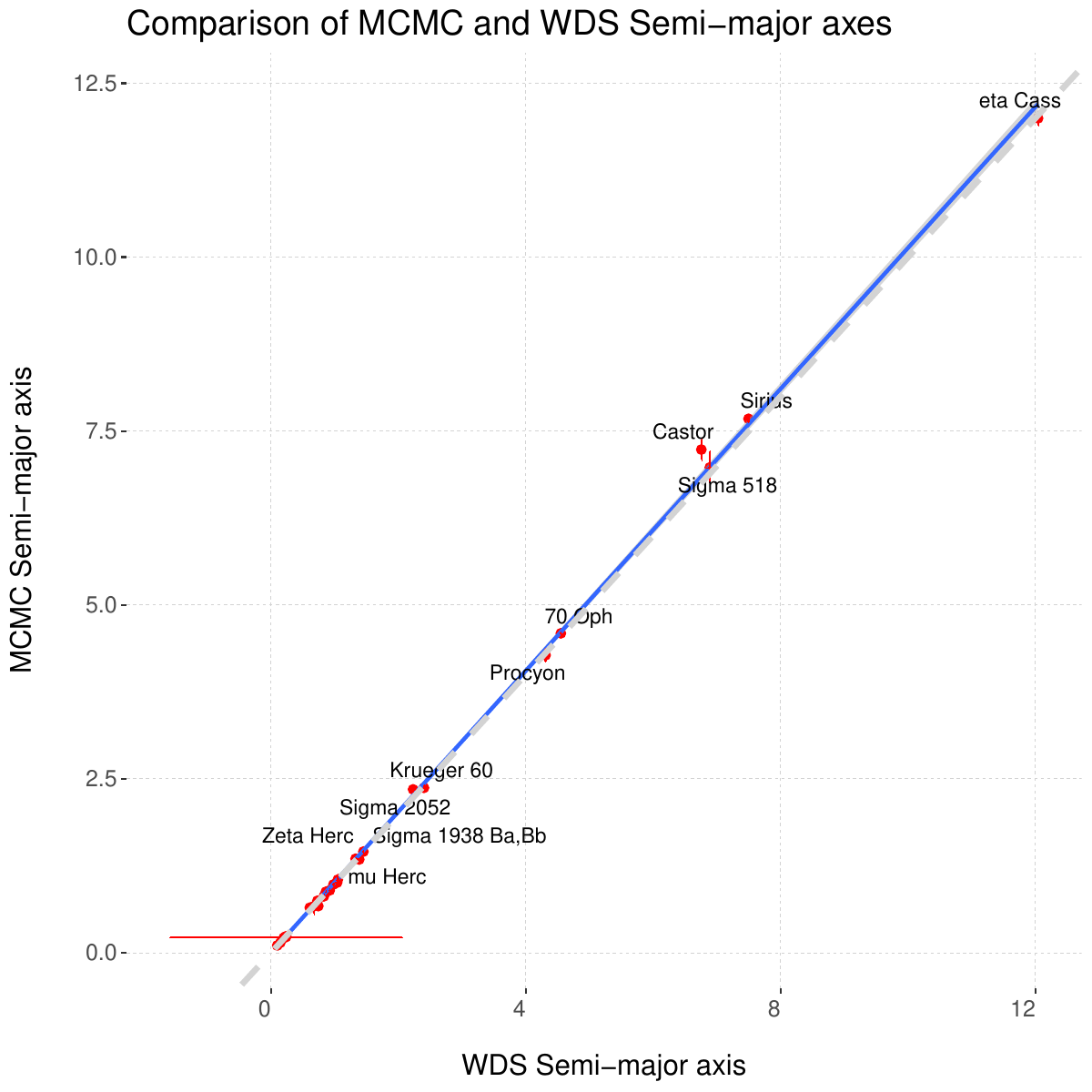}
  \caption{$a$}
  \label{fig:axis_comp}
\end{subfigure}\hfil 
\begin{subfigure}{0.45\textwidth}
  \includegraphics[width=\linewidth]{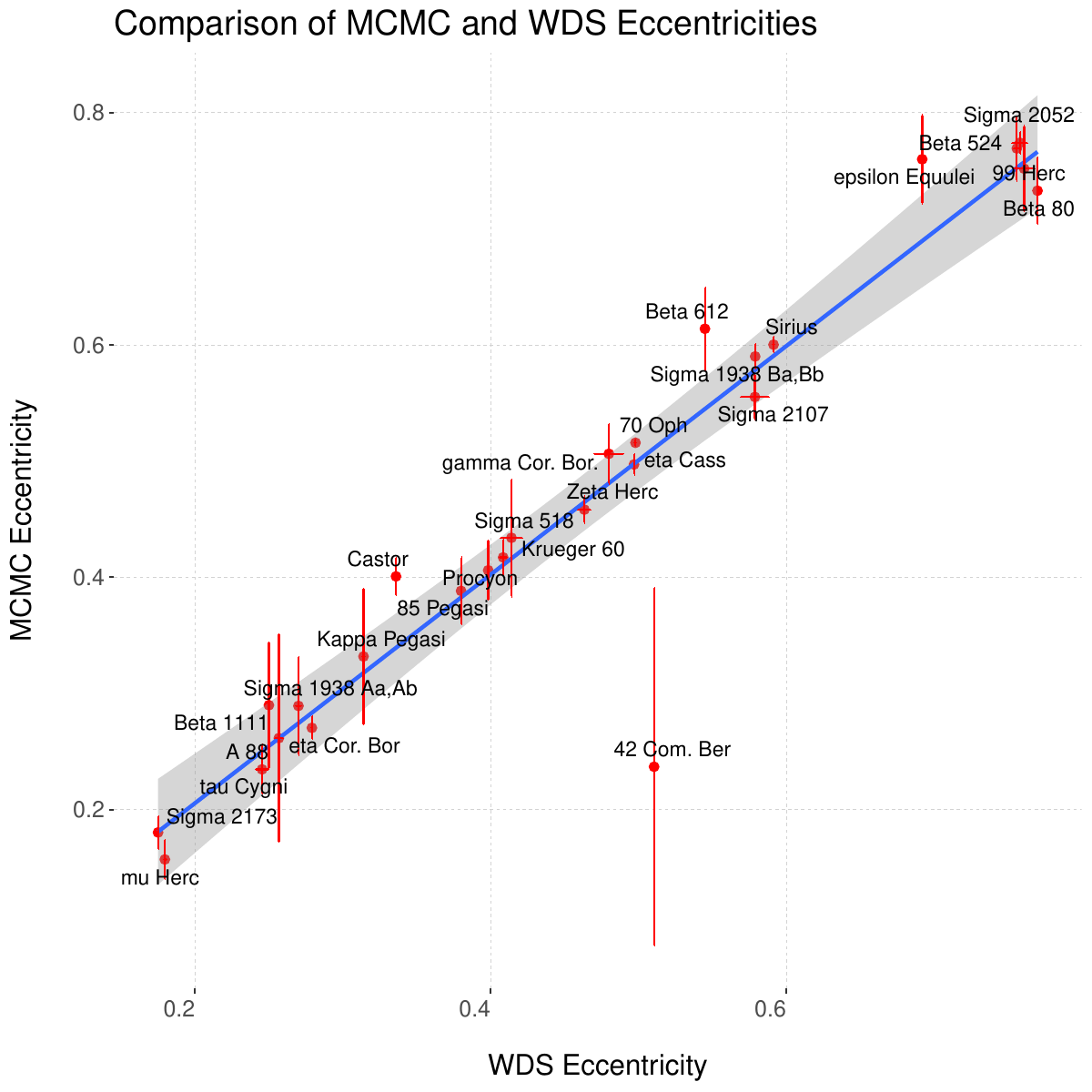}
  \caption{$e$}
  \label{fig:ellip_comp}
 \end{subfigure}
\begin{subfigure}{0.45\textwidth}
  \includegraphics[width=\linewidth]{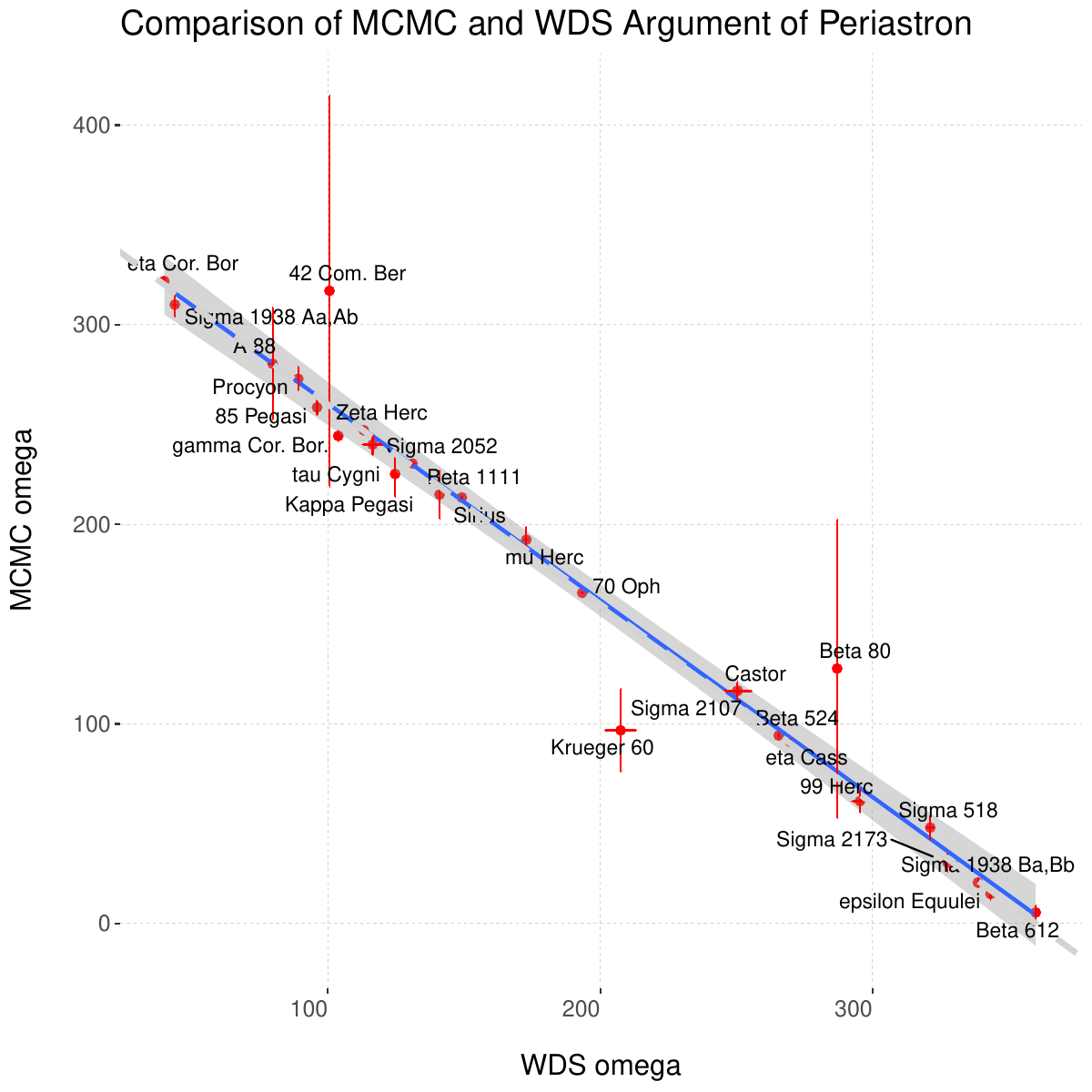}
  \caption{$\omega$}
  \label{fig:small_omega_comp}
\end{subfigure}
\caption{Comparison between MCMC (this paper) and WDS optimal parameter estimates for $P$, $a$, $e$ and $\omega$ by system. Linear regressions have been fitted to the data, resulting in best-fit (blue-colored) lines in the charts. Two sigma confidence limits are shown as the grey shaded regions. The dashed grey lines are those of perfect agreement, which are essentially in agreement with the regression lines indicating good agreement between the two sets of parameter estimates.
\label{fig:hmc_wds_comp}}
\end{figure*}

\begin{figure*}
    \centering 
\begin{subfigure}{0.48\textwidth}
  \includegraphics[width=\linewidth]{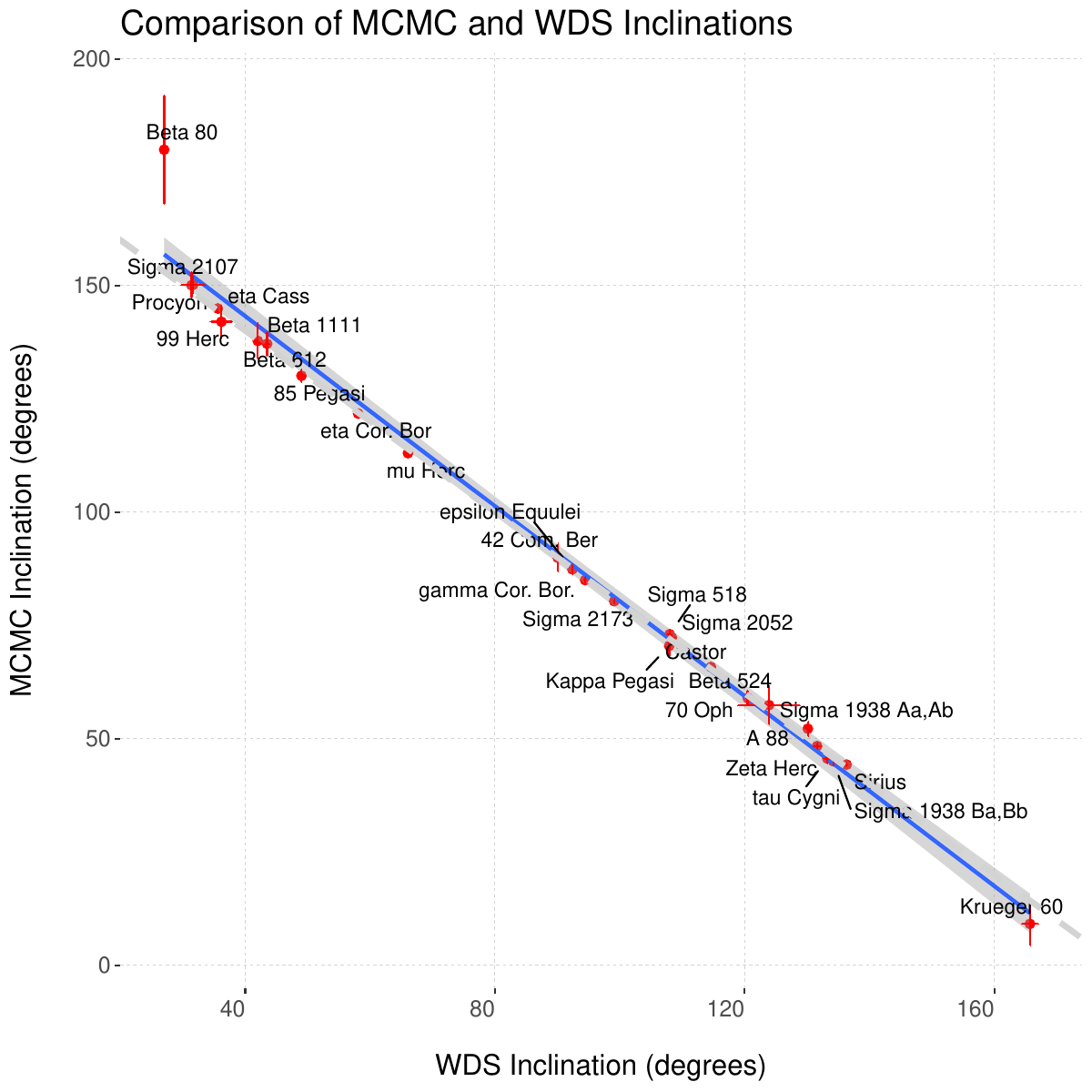}
  \caption{$i$}
  \label{fig:inc_comp}
 \end{subfigure}
\begin{subfigure}{0.48\textwidth}
  \includegraphics[width=\linewidth]{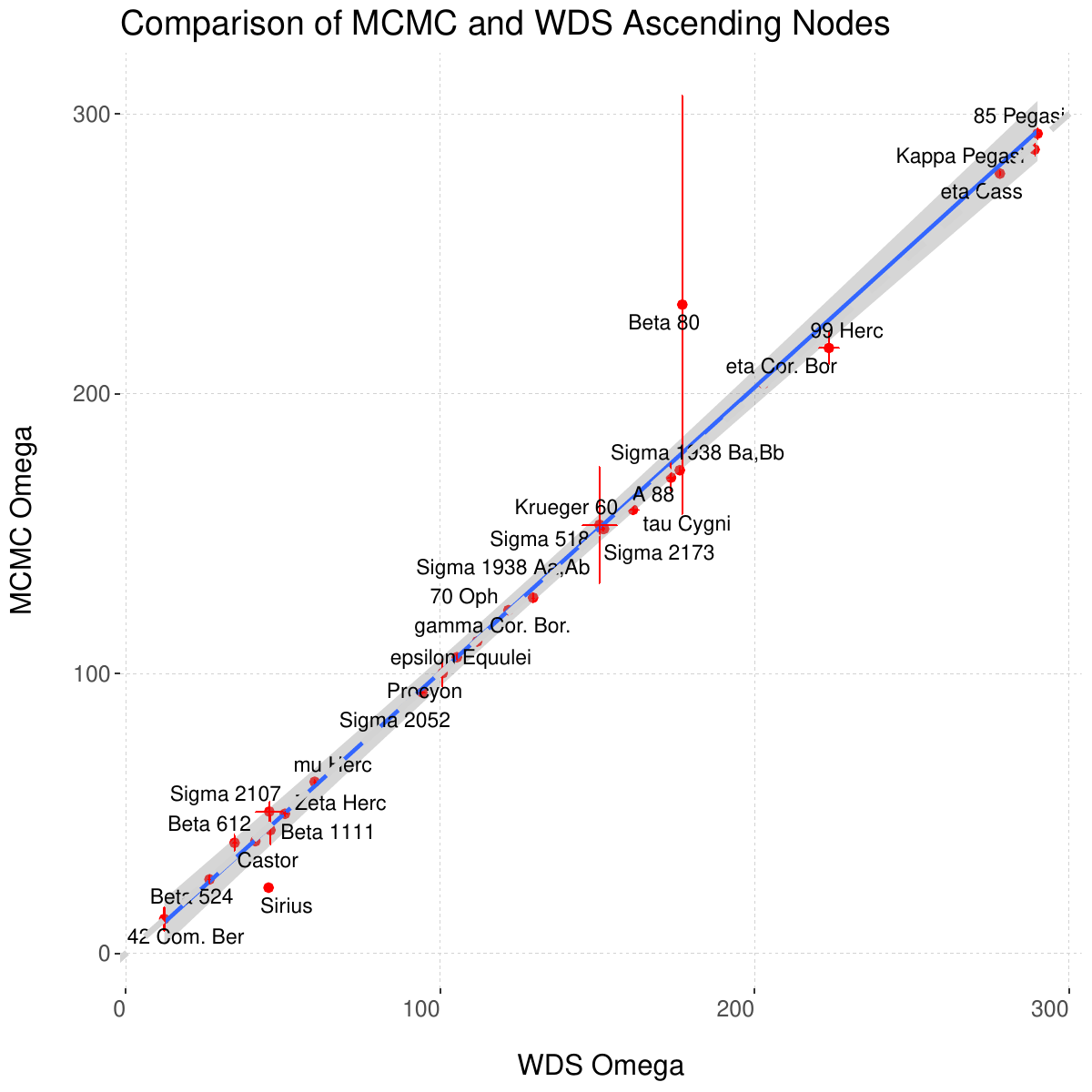}
  \caption{$\Omega$}
  \label{fig:big_omega_comp}
\end{subfigure}
\caption{Comparison between HMC and WDS optimal parameter estimates for $i$ and $\Omega$ by system.  
\label{fig:hmc_wds_comp2}}
\end{figure*}

\begin{table*}[!t]
    {\renewcommand\arraystretch{1.3}
    \caption{Dynamical solar masses ($M_D$) based on {\em Hipparcos} parallaxes (given in the column `Hipparcos' as milli-arcsconds) and parameter estimates from the MCMC orbital fitting. Errors in $M_D$ are one standard deviation. Column $M_M$ lists the dynamical masses given in \protect\cite{Malkov_2012}. $M_{VL}$ lists the dynamical mass estimates using the \protect\cite{vanLeeuwen_2007} parallaxes (listed in column `van Leeuwen' and in units of milli-arcseconds).
    \label{tab:dynamical masses}}
        \begin{center}
            \begin{tabular}{lcrlll}
\hline
Star            & $M_D$                     & $M_M$         & Hipparcos             &  van Leeuwen              & $M_{VL}$                      \\ 
\hline
42 Com  Ber     & $1.244_{-1.060}^{+2.731}$ &  ---          &   $ 69.81 \pm 27.58 $ &  $ 56.10 \pm 0.89 $       & $2.398^{+3.856}_{-1.478} $    \\ 
70 Oph          & $1.634_{-0.044}^{+0.046}$ &  1.60         &   $ 196.62 \pm 1.38 $ &  $ 196.72 \pm 0.83 $      & $1.631^{+0.031}_{-0.031} $    \\
85 Pegasi       & $1.474_{-0.241}^{+0.293}$ &  1.49         &   $ 80.63 \pm 3.03 $  &  $ 82.17 \pm 2.23 $       & $1.393^{+0.223}_{-0.192} $    \\
99 Herc         & $1.421_{-0.308}^{+0.368}$ &  1.73         &   $ 63.88 \pm 0.55 $  &  $ 63.93 \pm 0.34 $       & $1.418^{+0.349}_{-0.296} $    \\
A 88            & $2.540_{-0.716}^{+0.984}$ &  2.40         &   $ 20.01 \pm 0.93 $  &  $ 20.85 \pm 0.91 $       & $2.245^{+0.842}_{-0.620} $    \\
$\beta$  80     & $1.100_{-0.276}^{+0.377}$ &  0.43         &   $ 30.75 \pm 1.74 $  &  $ 31.20 \pm 1.60 $       & $1.055^{+0.337}_{-0.253} $    \\
$\beta$  524    & $4.461_{-1.395}^{+2.040}$ &  3.88         &   $ 13.87 \pm 0.86$   &  $ 14.15 \pm 0.72 $       & $4.201^{+1.709}_{-1.221} $    \\
$\beta$  612    & $2.991_{-0.618}^{+0.797}$ &  3.41         &   $ 18.68 \pm 0.91 $  &  $ 16.67 \pm 0.58 $       & $4.208^{+0.895}_{-0.733} $    \\
$\beta$  1111   & $0.746_{-0.312}^{+0.623}$ &  2.47         &   $ 22.40 \pm 3.40 $  &  $ 15.17 \pm 0.53 $       & $2.401^{+0.591}_{-0.475} $    \\
Castor          & $5.448_{-0.720}^{+0.838}$ &  5.43         &   $ 63.27 \pm 0.23 $  &  $ 64.12 \pm 3.75 $       & $5.234^{+1.923}_{-1.362} $    \\
$\epsilon$  Equ & $5.652_{-2.750}^{+6.831}$ &  4.17         &   $ 16.59 \pm 3.40 $  &  $ 18.49 \pm 1.35 $       & $4.082^{+1.607}_{-1.114} $    \\
$\eta$ Cass     & $1.567_{-0.106}^{+0.114}$ &  1.58         &   $167.99 \pm 0.62$   &  $ 167.98 \pm 0.48 $      & $1.567^{+0.110}_{-0.102} $    \\
$\eta$ Cor. Bor & $1.153_{-0.191}^{+0.241}$ &  2.11         &   $ 69.7 \pm 3.8 $    &  $ 55.98 \pm 0.78 $       & $2.229^{+0.147}_{-0.137} $    \\
$\gamma$ Cor Bor& $4.193_{-0.622}^{+0.738}$ &  4.18         &   $ 22.48 \pm 0.67 $  &  $ 22.33 \pm 0.50 $       & $4.278^{+0.640}_{-0.555} $    \\
$\kappa$ Peg    & $3.804_{-0.711}^{+0.871}$ &  ---          &   $ 28.34 \pm 0.88 $  &  $ 29.22 \pm 0.74 $       & $3.470^{+0.721}_{-0.661} $    \\
Krueger 60      & $0.586_{-0.179}^{+0.290}$ &  1.44         &   $ 225.0 \pm 25.6 $  &  $ 249.94 \pm 1.87 $      & $0.427^{+0.028}_{-0.026} $    \\
$\mu$ Herc      & $0.774_{-0.034}^{+0.036}$ &  ---          &   $ 119.05 \pm 0.62 $ &  $ 120.33 \pm 0.16 $      & $0.750^{+0.025}_{-0.025} $    \\
Procyon         & $2.014_{-0.153}^{+0.163}$ &  2.03         &   $ 285.93 \pm 0.88 $ &  $ 284.56 \pm 1.26 $      & $2.043^{+0.174}_{-0.162} $    \\
$\Sigma$ 518    & $0.855_{-0.141}^{+0.169}$ &  ---          &   $ 198.24 \pm 0.84$  &  $ 200.62 \pm 0.23 $      & $0.825^{+0.154}_{-0.130} $    \\
$\Sigma$ 1938   & $4.121_{-0.805}^{+0.976}$ &  ---          &   $ 26.96 \pm 0.65 $  &  $ 28.83 \pm 0.74 $       & $3.370^{+0.818}_{-0.671} $    \\
$\Sigma$ 2052   & $1.708_{-0.304}^{+0.376}$ &  1.63         &   $ 51.2 \pm 1.49 $   &  $ 50.87 \pm 0.80 $       & $1.742^{+0.298}_{-0.253} $    \\
$\Sigma$ 2107   & $2.630_{-0.582}^{+0.780}$ &  2.70         &   $ 17.62 \pm 0.95 $  &  $ 17.12 \pm 0.53 $       & $2.867^{+0.592}_{-0.482} $    \\
$\Sigma$ 2173   & $1.949_{-0.178}^{+0.199}$ &  1.91         &   $ 60.80 \pm 1.42 $  &  $ 61.19 \pm 0.68 $       & $1.912^{+0.118}_{-0.111} $    \\
Sirius          & $3.306_{-0.112}^{+0.115}$ &  3.08         &   $ 379.21 \pm 1.58 $ &  $ 379.21 \pm 1.58 $      & $3.306^{+0.115}_{-0.112} $    \\
$\tau$  Cyg     & $2.720_{-0.261}^{+0.289}$ &  ---          &   $ 47.80 \pm 0.61 $  &  $ 49.16 \pm 0.40 $       & $2.507^{+0.226}_{-0.209} $    \\
$\zeta$  Herc   & $2.619_{-0.124}^{+0.129}$ &  2.44         &   $ 92.63 \pm 0.60 $  &  $ 93.32 \pm 0.47 $       & $2.561^{+0.115}_{-0.110} $    \\
\hline
        \end{tabular}
    \end{center}}
\end{table*}  

We implemented this model as the fitting function in the {\sc stan} programming language,\footnote{For further details on {\sc STAN} see https://github.com/stan-dev/stan and https://mc-stan.org/users/documentation/} using the NUTS MCMC variant \citep{Hoffman_2014} to perform the optimisation. We note that if we were only interested in point estimates for the parameters, there are superior optimisation techniques which can reach such estimates with less computational effort.  Our key interest in using MCMC was to see how well constrained the parameter estimates are, rather than just the optimal estimates alone. The {\sc stan} code was called from the R programming language \citep{R_Core_2021}, where we handled data processing and additional analysis.  The role of MCMC was to adjust the model parameters so that the predicted positions became close to the actual data.  In other words, the optimizer trialed different estimates for the parameters in the model function, measuring how well the model based on this function fitted the observed data.  The measure of fit employed the $\chi^2$ function (see \citeauthor{Bevington_1969}, \citeyear{Bevington_1969}). The minimum chain length was 100,000 steps, with four chains being run simultaneously.  Convergence about an optimal solution set was assessed through trace plots (charts plotting parameter estimates by step position along the chains), which should be statistically random about the optimal estimates, i.e., no trends should remain.  We also made use of the \^{R} statistic \citep{Sinharay_2003} to assess convergence.

Best-fit solutions (and one standard deviation uncertainties) are listed in Table~\ref{tab:system_parameters} for each of the modelled systems. Figure~\ref{fig:orbit_plots} (page \pageref{fig:orbit_plots}) plots data for two example systems, along with the best-fit projected orbits based on the parameters given in Table~\ref{tab:system_parameters}. 

The orbital solutions are generally in good agreement with those listed by WDS as the adopted solutions for that catalogue, with the advantage that uncertainties for the parameters are given for all solutions.  Not all WDS solutions have uncertainties provided for the parameter estimates, as can be seen in Figures~\ref{fig:hmc_wds_comp} and \ref{fig:hmc_wds_comp2}. The NUTS-based uncertainties are generally larger than those given for the WDS solutions, even with our naive handling of errors. Eccentricity has the highest relative uncertainty out of the optimised parameters, followed by the argument of periastron.  42 Com Ber has an inclination close to 90 degrees and was difficult to model, leading to large uncertainties in parameter estimates for that system.

\begin{figure*}[!t]
\centering 
\begin{subfigure}{0.48\textwidth}
  \includegraphics[width=\linewidth]{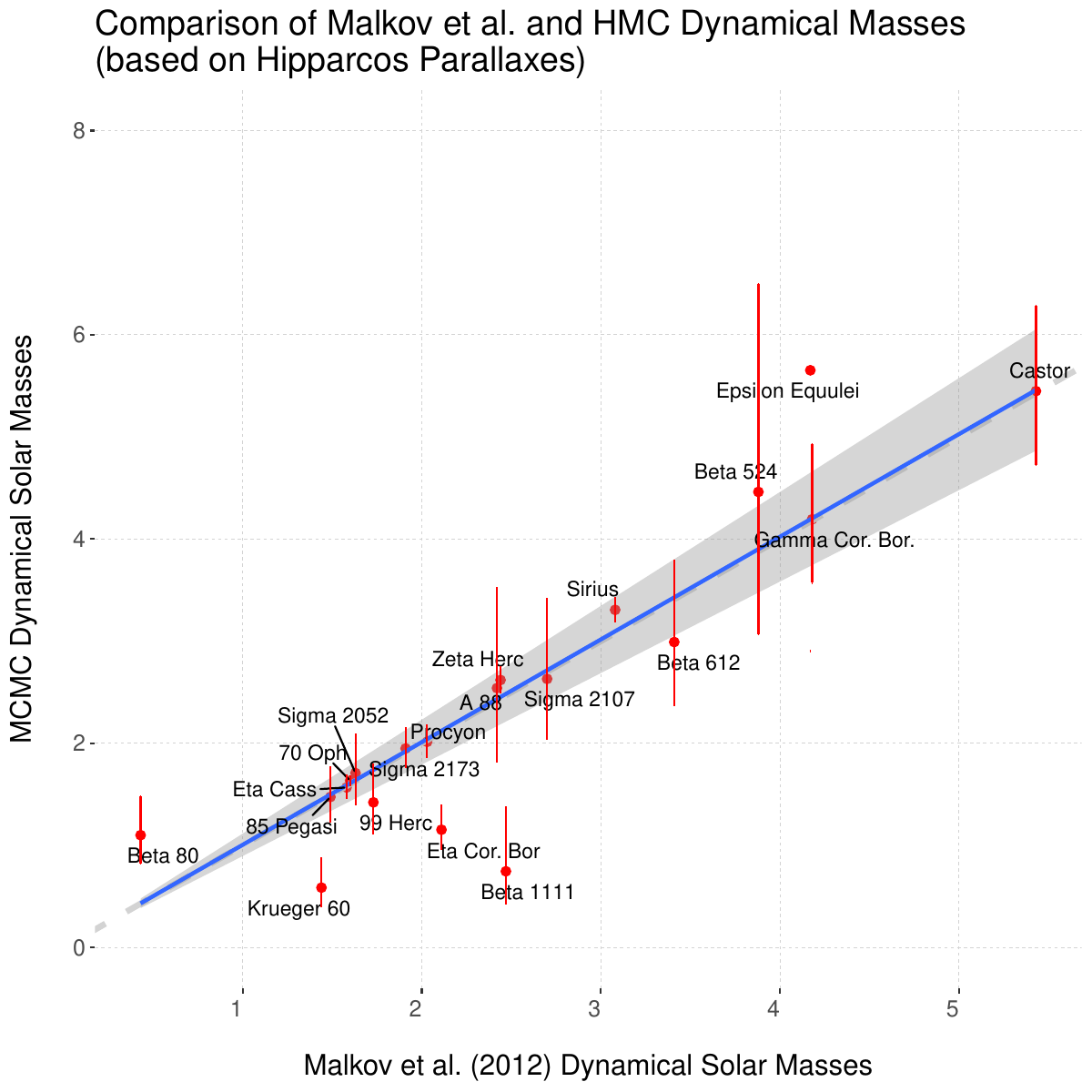}
  \caption{Using {\em Hipparcos} original parallaxes}
  \label{fig:hipparcos_masses}
 \end{subfigure}
\begin{subfigure}{0.48\textwidth}
  \includegraphics[width=\linewidth]{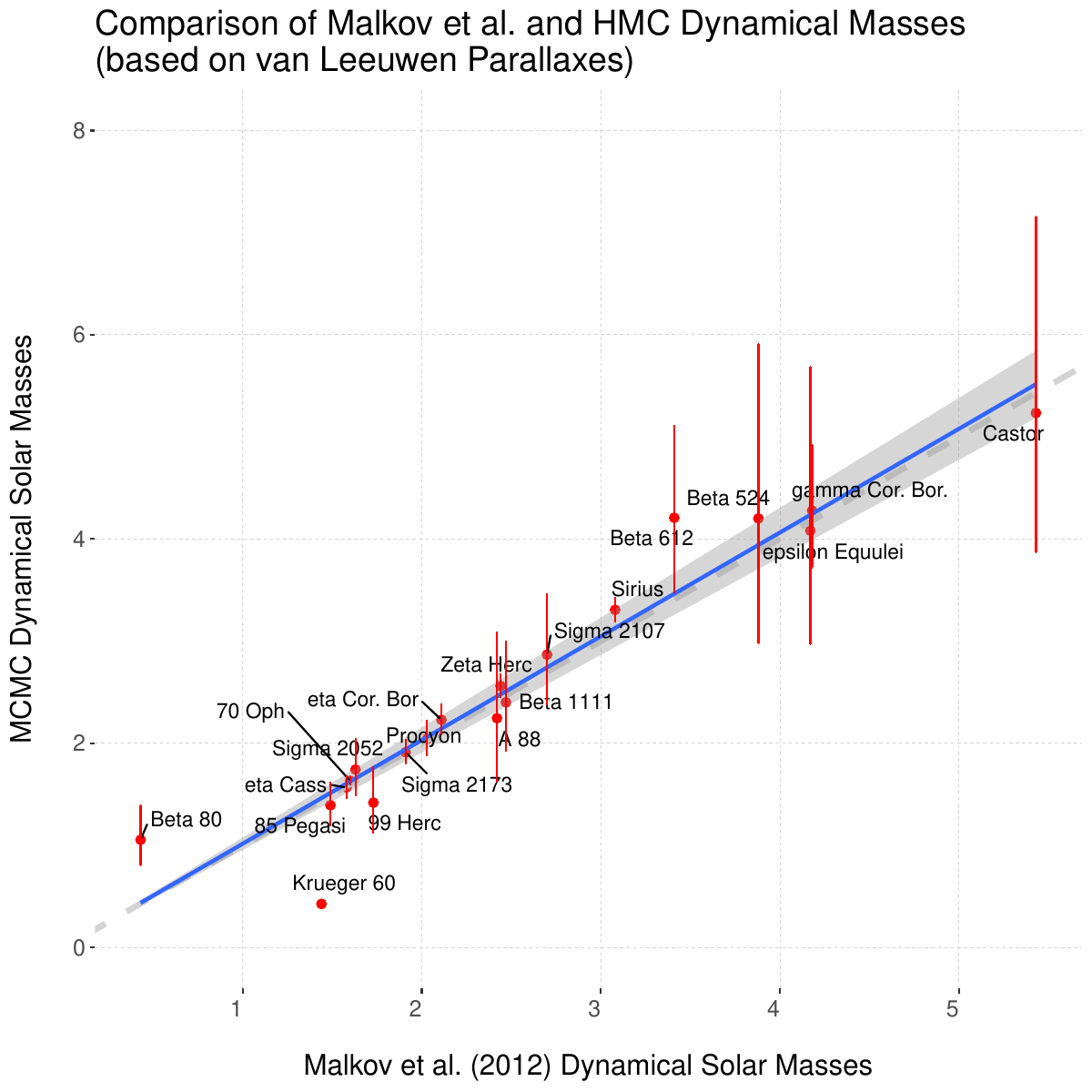}
  \caption{Using van Leeuwen parallaxes}
  \label{fig:vl_masses}
\end{subfigure}
\caption{A comparison of dynamical masses calculated in this study.  Sub-figure~\ref{fig:hipparcos_masses} is based on the {\em Hipparcos} parallaxes while sub-figure~\ref{fig:vl_masses} is based on \protect\cite{vanLeeuwen_2007}, as per equation~\ref{eq:masses}. These estimates are compared with those made by \protect\cite{Malkov_2012}, who did not provide uncertainties for their values. Overall agreement is good. Error bars are not plotted for Epsilon Equilei in sub-figure~\ref{fig:hipparcos_masses} as they are large and lead to a vertical compression of the chart when they are included, obscuring details. The shaded region corresponds to 95\% statistical confidence for the regression. A zero intercept was assumed. Note the improved scatter in sub-figure~\ref{fig:vl_masses} is largely driven by changes in the mass estimates for $\epsilon$ Equ, $\beta$ 1111, and $\eta$ Cor Bor. See Table~\ref{tab:dynamical masses} for the data plotted in these figures.
\label{fig:mp_parallaxes}}
\end{figure*}

The dynamical (or combined stellar) mass $M_d$ of such binary systems can be calculated if the parallax is known, via an equation \citep{Malkov_2012} based on Kepler's third law:
\begin{equation}
    M_{d} = \frac{a^3}{\pi^3 P^2}
    \label{eq:masses}
\end{equation}
where both $a$ and the parallax $\pi$ are in milli-arcseconds, $P$ is in years, and $M_d$ is in solar masses. 13 of the systems had {\em Gaia} DR3 parallaxes, and all 27 had {\em Hipparcos} parallaxes. The {\em Hipparcos} \citep{ESA_1997, Perryman_2008} and {\em Gaia} \citep{Gaia_2016, Gaia_2022} parallaxes agreed well (regressing the {\em Hipparcos} parallaxes onto the corresponding {\em Gaia} values resulted in a slope of $0.998 \pm 0.007$ assuming a zero intercept).  We therefore used the {\em Hipparcos} parallaxes for comparison given the good correlation and also the fact that more systems had parallax estimates in the {\em Hipparcos} dataset than the {\em Gaia} one.  We compared our estimates for the dynamical masses (Table~\ref{tab:dynamical masses}) with those from \cite{Malkov_2012}, see Figure~\ref{fig:mp_parallaxes}.  We found good agreement, indicating that our methodology appears reasonable, with the advantage that confidence intervals are generated for the optimized parameters.  However we also note that \cite{Malkov_2012} made use of the reduction by \cite{vanLeeuwen_2007} of {\em Hipparcos} astrometric data which improved parallax accuracies by up to a factor of four times for stars brighter than $\rm H_p = 8$, as well as the later analyses of \citep{Al-Wardat_2021} and \cite{Masda_2023} which demonstrated that the \citeauthor{vanLeeuwen_2007} parallaxes were superior to both the original {\em Hipparcos} and the DR3 {\em Gaia} estimates. Calculating dynamical masses using the \citeauthor{vanLeeuwen_2007}  parallaxes (see Table~\ref{tab:dynamical masses}) led to an improvement in the Pearson correlation coefficient from  0.905 to 0.960 (for the masses calculated in the current paper compared with those from \citeauthor{Malkov_2012}, \citeyear{Malkov_2012}), in line with the comments by \cite{vanLeeuwen_2007},  \cite{Al-Wardat_2021}, and \cite{Masda_2023}. We therefore recommend using the mass estimates given in the column $M_{VL}$ of Table~\ref{tab:dynamical masses} as the final estimates of the dynamical masses for the studied systems.


\section{Discussion}

This paper presents in Table~\ref{tab:system_parameters} new estimates of the orbital elements and uncertainties for a selection of systems listed in \cite{Miller_1922}, based on MCMC optimisation. We also calculate dynamical masses using equation~\ref{eq:masses} plus original \citep{ESA_1997, Perryman_2008} and refined \citep{vanLeeuwen_2007} {\em Hipparcos} parallaxes (see Table~\ref{tab:dynamical masses}), which we show to be in good agreement with \cite{Malkov_2012}. Figure~\ref{fig:vl_masses} shows the best comparison between our results and those of \cite{Malkov_2012}. Comparison of the orbital elements is made between those adopted by the WDS and those derived by our MCMC method (see Figures~\ref{fig:hmc_wds_comp} and \ref{fig:hmc_wds_comp2}). The estimates agree well, with the NUTS-based uncertainties tending to be significantly larger than those in the WDS-adopted solutions (not all such solutions provide formal errors). This comparison of known systems gives us confidence that the HMC-based technique presented could be reliably applied to new systems without previously published solutions. Indeed, we have used an earlier version of this methodology as part of the analysis of a multiple star system \citep{Erdem_2022}, with the astrometric analysis complementing and extending the spectroscopic and photometric analyses. We intend to use this methodology as we extend our survey of detailed studies of multiple systems (such as \citeauthor{Erdem_2022}) and recommend it to other researchers interested in not only estimating the orbital parameters but gaining insight into the accuracy of these estimates. We also hope that the orbital parameters (and accompanying uncertainties) presented by this paper for the 27 systems involved in the testing will be of interest to double star researchers, and that the paper acts as a record of the careful testing made of the methodology before its use for systems with no published estimates for orbital parameters or dynamical masses.


\section*{Acknowledgements}
This research has made use of the Washington Double Star (WDS) Catalog maintained at the U.S. Naval Observatory. We thank Dr.\ Rachel Matson for extracting data from the WDS for us. This work has made use of data from the European Space Agency (ESA) mission {\it Gaia} (\url{https://www.cosmos.esa.int/gaia}), processed by the {\it Gaia} Data Processing and Analysis Consortium (DPAC, \url{https://www.cosmos.esa.int/web/gaia/dpac/consortium}). Funding for the DPAC has been provided by national institutions, in particular the institutions participating in the {\it Gaia} Multilateral Agreement. We thank the University of Queensland for collaboration software. We thank the anonymous referee for their helpful comments and guidance which improved this paper.

\vspace{-1em}




\end{document}